\documentstyle[preprint,eqsecnum,aps]{revtex}
\begin{document}

\draft
\preprint{}
\title{Conservation laws for systems of extended bodies \\
 in the first post-Newtonian approximation}
\author{Thibault Damour}
\address{Institut des Hautes Etudes Scientifiques, 91440
 Bures sur Yvette, France \\
 and D\'epartement d'Astrophysique Relativiste et de
 Cosmologie, Observatoire de Paris, \\
 Centre National de la Recherche Scientifique, 92195
 Meudon CEDEX, France}
\author{David Vokrouhlick\'y\cite{byline1}}
\address{Observatoire de la C\^ote d'Azur, D\'epartement
 CERGA, Avenue N. Copernic, \\ 06130 Grasse, France}
\date{March 21, 1995}
\maketitle
\begin{abstract}
  The general form of the global conservation laws for $N$-body
  systems in the first
  post-Newtonian approximation of general relativity is
  considered. Our approach applies to the motion of an isolated
  system of
  $N$ arbitrarily composed and shaped, weakly self-gravitating,
  rotating,
  deformable bodies and uses a framework recently introduced by
  Damour, Soffel and Xu (DSX). We succeed in showing that seven of
  the first
  integrals of the system (total mass-energy, total dipole mass
  moment and total linear momentum)
  can be broken up into a sum of contributions which can be entirely
  expressed in terms of the basic quantities entering the DSX
  framework:
  namely, the relativistic individual multipole moments of the
  bodies,
  the relativistic tidal moments experienced by each body, and the
  positions and orientations with respect to the global coordinate
  system
  of the local reference frames attached to each body.
  On the other hand, the total angular momentum of the
  system does not seem to be expressible in such a form due to the
  unavoidable presence of irreducible nonlinear gravitational effects.
\end{abstract}

\narrowtext

\section{Introduction}

Recently, Damour, Soffel and Xu \cite{dam1,dam2,dam3,dam4} (DSX)
developed a new, exhaustive approach to the first post-Newtonian
dynamics of a system of $N$ extended bodies. This
theory is based on the complementary use of some local coordinate
systems (attached
to each body) and of a global coordinate system used to describe
the orbital motion of the $N$ bodies.
Detailed analyses of the laws of global translational motion
\cite{dam2} and local rotational motion \cite{dam3} of the bodies
have been given.

In this paper, we address the question of the global conservation laws
of an $N$-body system, and their link to the quantities introduced
in the DSX framework. We use the word conservation laws to mean the
first integrals related to the total four momentum and
angular momentum of the system, and to the center-of-mass theorem. The
existence, on the first post-Newtonian level, of these ten integrals
is guaranteed by the general form of the field equations
\cite{ll,fo,cha1,cha2,wei,mtw}.
Our main problem concerns {\it the form} of these conservation laws.
More precisely, we investigate whether they can be entirely expressed
in terms of the basic quantities introduced in the DSX post-Newtonian
theory: namely, the set of the relativistic individual
mass and spin multipole moments of the $N$ bodies
$(M_L^A,S_L^A)$ and the set of the gravitoelectric and gravitomagnetic
tidal moments experienced by each body $(G_L^A,H_L^A)$. [Here, as in
Refs.~\cite{dam1,dam2,dam3,dam4} whose notation we follow,
$A,B = 1, ..., N$ labels the various bodies, and $L = i_1 ... i_l$ is
a multi-spatial index] The former
fully characterize the structure of the post-Newtonian gravitational
field generated by an
extended body in its local coordinate system, while the latter
characterize the tidal action of the other bodies in this local
coordinate system, including inertial contributions due to its
acceleration and rotation. Obviously, the positions
and velocities, with respect to the global coordinate system, of the
origins of the local systems attached to each body, as well as their
orientations, need to be considered, and will also enter the final
expressions.

Let us start by discussing the various methods used for deriving some
explicit forms of the global conservation laws in general relativity.
First, let us remark that the expressions based on surface integrals
at infinity \cite{ll}, \cite{wei}, \cite{mtw} are of no real use
within the post-Newtonian
context because one loses a power $1/c^2$ in reading out a conserved
quantity from the asymptotic behaviour of metric. Concerning
approaches where the conserved quantities are given as
three-dimensional integrals, we note the formulation of Fock \cite{fo},
taken
up by Brumberg \cite{bru}, and that of Chandrasekhar and coworkers
\cite{cha1}, \cite{cha2}, followed by the parametrized version of
Will \cite{wi}. Neither
approaches constitute useful starting points for us. Indeed, on the
one hand, Fock and Brumberg restricted themselves to the
(ill-defined) case of a ``rigidly rotating'' bodies and introduced
some mass moments which are not compatible with the ones that
enter naturally the DSX formalism, while, on the other hand,
Chandrasekhar et~al. restricted themselves to the special case of
perfect fluids and chose basic variables which do not fit well
within the DSX scheme.

A more convenient starting point for our purpose is the work of
Blanchet, Damour and Iyer \cite{bd}, \cite{di1} (see also \cite{di2}).
These authors
have defined global-frame post-Newtonian mass and spin moments of an
arbitrary, isolated system of bodies in the form
of integrals over the compact supports of the bodies, and, they
explicitly checked that the lowest moments were conserved quantities.
Our task here will consist in transforming their
expressions, involving integrals over the global-time simultaneity
surface, into a combination of terms involving integrals over $N$
separate local-time simultaneity surfaces. Notice that nothing
guarantees a priori that these manipulations will lead to final
expressions containing only the ``good'' moments introduced in the
DSX scheme. In fact, we shall succeed in this task for seven of the
globally conserved quantities and fail for three of them (the three
components of the trickier total angular momentum).

Let us note in advance that there are several limiting cases for
which it is already known that some of the globally conserved
quantities can be entirely expressed in terms of some dynamically
relevant individual multipole moments. First, there are the cases
where one truncates the multipolar series: keeping only the monopole
contributions defines the ``Lorentz-Droste-Einstein-Infeld-Hoffmann''
model (LD-EIH), while keeping also the intrinsic spins of the bodies
define the ``pole-dipole'' model (PD). In both cases, we dispose
of well established forms of the conservation laws (see e.g.
\cite{ll},
\cite{fo} for the monopole case, and \cite{dam}, \cite{ds} for
the pole-dipole case).
These will provide useful checks on our general results. Another
limiting situation of possible relevance is that of a test
point-like mass moving in the gravitational
field of $N-1$ extended bodies. The motion of the
artificial/natural solar system satellites is a typical case of this
category.  Ref.~\cite{dam4} computed explicitly the form of the
corresponding Lagrangian in terms of the gravitational potentials
$(W,W_a)$, which can be algorithmically constructed using the
formulas given in the Appendix of \cite{dam2}. However, in this case
we have first integrals only when the Lagrangian (describing geodesic
motion) possesses some continuous symmetries (Noether's theorem).
For instance, we can consider $N=2$ with $M_2 << M_1$ (restricted
two body problem) and
with a stationary and/or axially symmetric central body (see e.g.
Ref.~\cite{heim} for a study of a restricted problem of this type).
This type of limiting situation will not give us useful checks.

Finally, let us remark that we hope that the present work will find
practical applications in the relativistic motion of binary stars
or in the celestial mechanics of the solar system. Let us recall,
for instance, that one simplifies
the dynamical ephemeris of the solar system (at the Newtonian
approximation) by eliminating the motion of the Sun via the algebraic
relation expressing that the total dipole mass moment of the solar
system vanishes for all time (in a suitably mass-centered frame;
\cite{jpl}).

In Sec.~II we consecutively discuss the cases of the mass-energy
integral, linear momentum integral and briefly comment on the
angular momentum integral. Sec.~III contains a summary of our results.
Throughout the paper we follow the terminology and notation used
in Refs.~\cite{dam1,dam2,dam3,dam4}.

\section{FIRST INTEGRALS OF THE DYNAMICAL LAWS}

\subsection{Comments on the change of time-simultaneity
 integration domains}

As mentioned in the preceding section, transformation of integrals
{}from the global to several local time simultaneity surfaces is
a common point to all particular cases of conservation laws.
We shall thus start our discussion with a brief technical comment
concerning such transformations.

Considering a specific body $A$, we study some integral
\begin{equation}
 {\cal I}_A(t) = \int_A d^3x\, f(t,{\bf x})\; ,\label{one}
\end{equation}
performed on the global time $t={\rm const}.$ surface, where
$f(x^\mu)$ is some given function defined on the {\it compact}
support of body $A$. We seek a transformation of the right hand side
of ${\cal I}_A$
such that it can be written as an integral on a local-time
$T_A=T_A^0(t)={\rm const}.$ simultaneity surface, say
\begin{equation} {\cal I}_A(T_A^0(t)) = \int_A d^3X_A\, {\tilde F}
 (T_A^0(t), {\bf X}_A)\; .\label{two} \end{equation}
Here, $T_A^0(t)$ denotes the value of the local time
(in the reference
system attached to body $A$) corresponding to the event on the
{\it central} worldline of the body-$A$ reference system (${\bf X}_A
= 0$, also quoted as ${\cal L}_A$), whose global time coordinate is
equal to $t$. In equations, if we
write the spacetime coordinate transformation between the global
coordinate system $x^\mu=(ct,x^i)$ and the body-$A$ local one
$X_A^\alpha = (cT_A,X_A^a)$ as
\begin{eqnarray}
 x^\mu = x^\mu(X_A^\alpha)& =& z_A^\mu(T_A) + e^\mu_{A a}(T_A)
  \bigl[X^a_A \nonumber \\
 & & \;\; + {1 \over 2c^2} A^a_A {\bf X}_A^2 - {1 \over c^2}
 ({\bf A}_A. {\bf X}_A) X_A^a \bigr] \; ,\label{three}
\end{eqnarray}
$T_A^0(t)$ is defined as the unique solution of $ct =
x^0(T_A,{\bf 0})$, i.e. $ct \equiv z_A^0(T_A^0(t))$.

First, let us denote by $f_A(T_A, {\bf X}_A)$ the original function
$f(t,{\bf x})$ reexpressed in terms of the local spacetime variables
$X_A^\alpha$: $f_A(X_A^\alpha) \equiv f(x^\mu(X_A^\alpha))$. By
mathematically transforming the variables of integration in
Eq.\ (\ref{one}) we get
\begin{equation}
 {\cal I}_A(t) = \int_A d^3X_A \left( \left|{\partial X \over
 \partial x}\right|^{(3)}_{t = const}\right)^{-1}\!\!\! f_A\left[
 T_A\left(t, {\bf X}_A\right),{\bf X}_A\right] \; ,\label{four}
\end{equation}
where $|\partial X / \partial x|^{(3)}_{t = const}$ is the
spatial Jacobian ${\rm det}(\partial X^a / \partial x^i)$ $(a,i =
1,2,3)$ computed when keeping fixed the value of $t$, and where
$T_A(t,{\bf X}_A)$ denotes the solution of $ct = x^0(T_A,{\bf X}_A)$.
 From Eq.\ (\ref{three})
(or Eqs.~(A5) of Ref.~\cite{dam2}), the latter quantity reads
explicitly
\[ T_A(t,{\bf X}_A) = T_A^0(t) - {1 \over c^2} V_A^a X_A^a +
 {\cal O}(4)\; , \]
so that by expanding $f_A[T_A(t,{\bf X}_A), {\bf X}_A]$
in powers of the small time shift $({\bf V}_A.{\bf X}_A)/c^2 =
{\cal O}(2)$ we can express it within a sufficient accuracy in terms
of functions computed on a {\it local-time} simultaneity surface,
namely $T_A = {\rm const.} = T_A^0(t)$:
\begin{eqnarray}
 f_A\left[T_A\left(t,{\bf X}_A\right),{\bf X}_A\right] &=&
 f_A\left(T_A^0(t),{\bf X}_A\right) \nonumber \\
 & & - {1 \over c^2} V_A^a X_A^a
 \partial_T f_A\left(T_A^0(t),{\bf X}_A\right) \nonumber \\
 & & + {\cal O}(4) \; .  \label{five}
\end{eqnarray}

As for the three-dimensional Jacobian entering Eq.\ (\ref{four}) it
is easy to see that it can be expressed as
\begin{eqnarray}
 \left|{\partial X \over \partial x} \right|^{(3)} &= &
 \left({\partial t \over \partial T_A}\right)_{X^a = const}
 \left|{\partial X \over \partial x} \right|^{(4)} \nonumber \\
 &=& \left[e^0_{A 0}(T_A^0) + {1 \over c^2} A^a_A X^a\right]
 \left|{\partial X \over \partial x} \right|^{(4)} + {\cal O}(4)
 \; .\label{six}
\end{eqnarray}
where $|\partial X / \partial x|^{(4)} = {\rm det}(\partial X^\alpha
/ \partial x^\mu)$ is the full four-dimensional Jacobian associated
with the coordinate transformation (\ref{three}). The time derivative
$(\partial t/\partial T_A)$ in Eq.\ (\ref{six}) is obtained from
Eq.\ (\ref{three}) or from
putting $V_S = 0$ in expression (A6) in Appendix A of \cite{dam4}.

Finally, we get
\begin{eqnarray}
 {\cal I}_A(t) &=& \int_A d^3X_A\, \left[ \left|{\partial X
 \over \partial x}\right|^{(4)}\right]^{-1}\! \left[e^0_{A 0}\left(
 T_A^0 \right) + {1 \over c^2} A_A^a X_A^a\right]^{-1} \nonumber \\
 & &\quad \times \left[f_A\left(T_A^0,{\bf X}_A\right) - {1 \over
 c^2} V_A^a X_A^a \partial_T f_A\left(T_A^0,{\bf X}_A\right)\right]
 \; , \nonumber \\ & & \label{seven}
\end{eqnarray}
where it is convenient to leave unexplicated the
four-dimensional Jacobian because it will be directly cancelled
when using the transformation laws of the mass and mass current
densities $\sigma^\mu$ entering $f_A(t,{\bf x})$. For completeness,
let us however mention its value
\begin{equation}
 \left| {\partial X \over \partial x} \right|^{(4)} = 1 -
 {2 \over c^2} W''(T,X) + {\cal O}(4) \; , \label{eight}
\end{equation}
where $W''(T,X) = G''_A - A_A^a X^a + {\cal O}(2)$ is the inertial
contribution to the local potential due to the change of the
time scale and the acceleration of the body
$A$ frame. Here, $ G''_A = c^2 \ln(dT / d\tau_f)_A = v_A^2/2 -
c^2 \ln e^0_{A 0} + {\cal O}(2)$ measures the relative scaling, along
the
central worldline $X_A^a = 0$, between the local time $T_A$ and the
global Minkowskian proper time $d\tau_f = \sqrt{-f_{\mu\nu}dz^\mu
dz^\nu}/c$ (see Sec.~VI.E of \cite{dam1}).

Note that, in geometrical terms, the mathematical transformations we
have just performed correspond to using a mapping between the
$t={\rm const.}$
and $T_A = {\rm const.} = T_A^0(t)$ hypersurfaces by means of the
congruence
of worldlines ${\cal L}_{X_A^a}$ of constant spatial local coordinates
(see Sec.~III.D in \cite{dam1}).

\subsection{Mass-energy integral}

The Blanchet-Damour post-Newtonian total mass-energy $m(t)$ of an
isolated system can be written as \cite{bd}
\begin{equation}
 m(t) = \int d^3x\, \sigma(x) - {1 \over c^2} {d \over dt} \int d^3x
  \,(\sigma^i x^i) + {\cal O}(4) \; , \label{nine}
\end{equation}
where $\sigma_\mu$ are densities of mass and mass currents
[$\sigma = (T^{00}+T^{ii})/c^2$ and $\sigma^i = T^{0i}/c$; $T^{\mu
\nu}$ denoting the components of the stress-energy tensor in the
{\it global} coordinate system].
The integration in (\ref{nine}) is to be performed over a global-time
$t={\rm const}.$ hypersurface spanning the whole $N$-body system.
However, as $\sigma^\mu$ is nonzero only in the neighbourhood of the
bodies, we can directly use the results of the previous sub-section
to decompose $m(t)$ as a sum of $N$ terms integrated over local
simultaneity surfaces $T_A^0(t) = {\rm const.}$.

Let us recall the transformation law between the global and local
coordinate systems of the mass densities pertaining to a given
point of body A
\begin{eqnarray}
 \sigma(x) & =& \left|{\partial X \over \partial x} \right|^{(4)}
  \Bigl[\left(1+2{v_A^2 \over c^2}\right)\Sigma(X) \nonumber \\
 & & \qquad\quad\quad + {4 \over c^2}
  V_A^a \Sigma^a_A(X)\Bigr] + {\cal O}(4) \; , \label{ten}
\end{eqnarray}
as given in \cite{dam1}.

Putting together the definition (\ref{nine}) and the method explained
in the previous sub-section II.A [where we note that the
four dimensional Jacobian cancels between Eqs.\ (\ref{seven}) and
(\ref{ten})] we arrive after some algebra to
\begin{eqnarray}
  m(t) &=& \sum_A \biggl[ M^A\left(1+{1 \over 2c^2}
   v_A^2\right) + {1 \over c^2} {d \over dT_A}\left(M^A_a V_A^a\right)
  \nonumber \\
  & & \qquad + {1 \over 2c^2} \sum_l {2l+1 \over l!} M^A_L
   \left(G_L^A + G_L^{A\prime\prime} \right)\biggr] + {\cal O}(4) \; ,
 \nonumber \\ \label{eleven}
\end{eqnarray}

\noindent where all quantities on the right-hand side must be
evaluated at
the intersection of the $t={\rm const.}$ hypersurface with the
central worldline
of the corresponding body (i.e. for $T_A = T_A^0(t)$). We recall that
the quantities $M^A$ (which are not constant in general) denote the
local, individual relativistic mass monopoles of each body. Each
$M^A$ is a directly observable quantity in the sense that it is just
the
gravitational mass measured from interpreting the locally measured
orbital motion of artificial/natural satellites around body $A$.
Similarly the $M^A_L$ (for $l \geq 1$) are the locally measured mass
multipole moments of body $A$, and $G^A_L$ the locally felt tidal
moments. The other quantities entering Eq.\ (\ref{eleven}) are related
to the way
the local $A$-reference system is moving with respect to the global
coordinate system [in particular $G''^A_L = (G''^A,-A^A_a, +3 A^A_{<a}
A^A_{b>}/c^2, 0, 0, ...)$ measure the inertial contributions to the
tidal moments felt in the local $A$ system]. The result (\ref{eleven})
is new, and its precise form (e.g. the numerical factor $2l+1$ in
front
of the $M^A_L G^A_L$ product) is different from what one might have
naively expected from the standard Newtonian expression for the total
energy (e.g. \cite{hart}).

As discussed in detail in Ref.~\cite{dam1} the DSX framework leaves
open some
freedom in fixing several quantities related to the origin an
orientation
of the local coordinate systems, as well as the gauge for the time
coordinate along the worldline ${\cal L}_A$. We shall call ``standard
worldline data'' the case where this freedom is used to satisfy
the following constraints: (i)
$\forall A\, ,\forall T_A\, ,M_a^A(T_A)=0$ (which means identifying
the origin of all local coordinate systems with the relativistic
mass centers of the bodies), (ii) $\forall A\, ,
\forall T_A\, , {\bar W}^A_\alpha(T_A,0)=0$ (the so called weak
effacement condition for the external gravitational potential in the
local frames). Note that we still leave unconstrained the orientation
of the
local frame axes which can undergo a slow rotation described
by the matrix $R^i_{A a}(T)$. In the case of the standard worldline
data, Eq.\ (\ref{eleven}) simplifies to the form
\begin{eqnarray}
  m^{\rm standard}(t) &=& \sum_A \biggl\{ M^A\left[1+{1 \over 2c^2}
   \left( v_A^2- G_{A}^{\prime}\right)\right] \nonumber \\
 & & + {1 \over 2c^2} \sum_{l \geq 2} {2l+1
   \over l!} M^A_L G_L^A  \biggr\} + {\cal O}(4) \; , \label{twelve}
\end{eqnarray}
where $G_{A}^{\prime} = \sum_{B \neq A} G^{B/A} = \sum_{B \neq
A} w^B(z_A) + {\cal O}(2)$ [$G_A' = -G_A''$ for standard data] denote
the value on the central worldline ${\cal L}_A$
of the Newtonian potential generated by all the other bodies [$
w^B(z_A) = \sum_{l \geq 0} {(-)^l \over l!} G M^B_L \partial^A_L
|z_A - z_B|^{-1}$].

In the monopole (LD-EIH) or pole-dipole limit Eq.\ (\ref{twelve})
yields the well-known result that the total mass-energy is
the sum of the total rest-mass and the kinetic and potential
($\propto GM^AM^B/|z_A-z_B|$) energy terms. Let us recall that, in
the general case, the individual post-Newtonian gravitational masses
$M^A$ are no longer constant (because of tidal forces acting on the
extended bodies, see Eqs.~(4.20a) and (4.21a) of \cite{dam2}) and that
it is not a priori evident that the quantities $m(t)$ or $m^{\rm
standard}(t)$ are conserved. As a check on our algebra, we have
verified by a direct calculation that this is indeed the case.

As an aside, let us conclude this sub-section by noting that if one
defines, as an auxiliary technical quantity, the ``Fock mass''
of the $A$-th body by the relation $c^2 M^A_F =
\int_A d^3X_A\, (1+W/2c^2) {\bf T}^{00}_A$ (where ${\bf T}^{00}_A$
denote the
$X_A^0$-$X_A^0$ component of the stress-energy tensor considered in
the local coordinate system $X_A^\alpha$), one can write our result
(\ref{twelve}) in a formally compact (and familiar looking) form.
The name we give to this
quantity is based on the fact that in his book \cite{fo} Fock used
such
an expression for the total mass-energy. Note, however, that he
always
used it in the global coordinate system. Our definition is written in
the local system $X_A^\alpha$, but we use for $W(T,X)$ the {\it total}
potential in the local frame, containing both internally and
externally
(including inertially) generated contributions. By using the
expressions (4.15) of \cite{dam2} for the tidal expansion of $W(T,X)$
we find (in any worldline gauge) the following relation between the
Blanchet-Damour mass and the Fock one:
\begin{equation}
 M^A  = M^A_F - {1 \over 2c^2} \sum_{l} {2l+1 \over l!} M^A_L
  G_L^A + {\cal O}(4) \; . \label{thirteen}
\end{equation}
Thus in the case of standard data we can rewrite our previous
formula (\ref{twelve}) in the following form
\begin{equation}
 m^{\rm standard}(t) = \sum_A M^A_F\left[1+{1 \over 2 c^2}\left(
   v_A^2- G_{A}^{\prime}\right)\right] + {\cal O}(4) \; .
 \label{fourteen}
\end{equation}
Let us, however, emphasize again that it is only the
Blanchet-Damour mass $M^A$ which is directly observable as a
gravitational mass determining the orbital motion of satellites
of body $A$. The Fock mass $M^A_F$ is just a mathematical construct.

\subsection{Center-of-mass integral and linear momentum}

The Blanchet-Damour post-Newtonian dipole mass moment $m_i(t)$ of
the whole system \cite{bd} can be written as
\begin{eqnarray}
 m_i(t) &=& \int d^3x \,x^i \sigma(x) \label{fifteen} \\
 & & \quad - {1 \over c^2} {d \over dt} \int
  d^3x \,\sigma^j \left(x^i x^j - {1 \over 2} \delta_{ij} x^2\right)
  + {\cal O}(4)  \; . \nonumber
\end{eqnarray}
It satisfies the following conservation law \cite{bd}
\begin{equation}
 {d^2 m_i(t) \over dt^2} = 0 \; . \label{sixteen}
\end{equation}
In fact, the first derivative of $m_i(t)$ is nothing but the
conserved total linear momentum of the system
\begin{equation}
 p_i \equiv {d m_i(t) \over dt} = {\rm const.}   \; .
 \label{seventeen}
\end{equation}

Employing the results of Sec.~II.A and definition (\ref{fifteen})
we obtain after tedious but straightforward calculations the following
form of the total mass dipole moment $m_i(t)$ of the system
\widetext
\begin{eqnarray}
 m_i(t) &=& \sum_A\biggl\{ M^A z_A^i \left[1+{1 \over c^2}\left(
  {1 \over 2}v_A^2+G_{A}^{\prime\prime} \right)\right] + {1 \over c^2}
  v_A^j s_A^{ij} - {3 \over c^2} a_A^j m_A^{ij} \nonumber \\
 & & \qquad\; - {1 \over c^2} \left(z_A^i z_A^j - {1 \over 2}
  \delta_{ij}
  z_A^2 \right) R^j_{A a} \sum_l {1 \over l!} M_L^A G_{aL}^{A\prime}
  - {2 \over c^2} z_A^j R^i_{A [a} R^j_{A b]} \sum_l
  {1 \over l!} M_{aL}^A  G_{bL}^{A\prime} \biggr\} \nonumber \\
 & & +  \sum_A \biggl\{m^i_A \left[1+{1 \over c^2}\left(
  {1 \over 2}v_A^2+G_{A}^{\prime\prime} \right)\right]
  + {1 \over c^2} z_A^i \left[M^{A (1)}_a V_A^a + 2 M^A_a
  G_a^{A\prime\prime} \right]\biggr\} \nonumber \\
 & & + {\cal O}(4) \;, \label{eighteen}
\end{eqnarray}
\narrowtext
\noindent where $s^{ij}_A  = \epsilon_{ijk} s^k_A = \epsilon_{ijk}
R^k_{A a}
S^a_A$, $m^{i_1i_2 ...i_n}_A = e^{i_1}_{A a_1}e^{i_2}_{A a_2} ...
e^{i_n}_{A a_n} M_A^{a_1a_2 ... a_n}$, where $M^{A(1)}_a = dM^A_a
/ dT_A$ and where the brackets in $R^i_{A [a}
R^j_{A b]}$ mean anti\-symme\-tri\-za\-tion [$u_{[a} v_{b]} \equiv
\frac{1}{2}(u_a v_b - u_b v_a)$]. It should be emphasized that the
fact that
the final expression (\ref{eighteen}) can be entirely written in terms
of the ``good'' moments entering the DSX framework is far from being a
trivial result. In the intermediary calculations the ``bad'' moments
$(N_L,P_L)$ defined in Eqs.~(4.22) of \cite{dam2} enter at several
places before finally cancelling. We also
remark that the last sum in curly brackets vanishes if one uses
standard worldline data, as defined above.

As mentioned in Sec.~I, we can get partial checks on our results by
considering models where the multipole series is highly truncated. In
particular, if we keep only the mass monopoles of the bodies (LD-EIH
limit), formula (\ref{eighteen}) reduces to
\begin{eqnarray}
 m^i_{LD-EIH} &=& \sum_A  M^A z_A^i \left[1+{1 \over 2c^2}
  \left(v_A^2 - G_A^\prime\right)\right] + {\cal O}(4) \nonumber \\
  &=& \sum_A  M^A z_A^i\left[1+{1 \over 2c^2}
  \left(v_A^2 - \sum_{B \neq A} {GM^B \over r_{AB}}\right)\right]
 \nonumber \\ & & \quad + {\cal O}(4) \; , \label{nineteen}
\end{eqnarray}
where the second row applies to the case of standard
worldline data. Eq.\ (\ref{nineteen}) agrees with previous results
\cite{ll}, \cite{mtw}. In the case of the pole-dipole truncated model
the expression for the mass dipole reads
\begin{equation}
 m^i_{PD} = m^i_{LD-EIH} + {1 \over c^2} \sum_A v^j_A s^{ij}_A +
 {\cal O}(4) \; , \label{twenty}
\end{equation}
a result previously derived by Damour and Sch\"afer \cite{ds} from
the spin dependent Lagrangian of Ref.~\cite{dam}.

Let us remark, as an aside, that defining some ``Fock'' local mass
dipole moments for instance by $c^2 M^A_{a F} =
\int_A d^3X_A\, X_A^a(1+W/2c^2){\bf T}^{00}$ does not simplify at all
the writing of our result (\ref{eighteen}). In fact, this definition
introduces
several bad algebraic structures (notably the moments $N^A_L$; see
\cite{dam1,dam2,dam3}) which do not enter the final dynamical results
of the DSX formalism. This is one of the reasons why Fock, in his
book \cite{fo},
did not succeed in getting a good definition of the mass centers of
the individual bodies (in spite of the fact that, in the case of the
mass center of the entire isolated system, the Blanchet-Damour and
Fock definitions, written in the global coordinate system, give the
same result; see Eq.~(3.45) of \cite{bd}).

{}From the result (\ref{seventeen}) above, we can easily derive the
following explicitly DSX-like expression for the total linear momentum
of the $N$-body system:
\widetext
\begin{eqnarray}
 p_i(t) &=& \sum_A\biggl\{ M^A v_A^i\left[1+{1 \over c^2}\left(
  {1 \over 2}v_A^2+G_A^{\prime\prime} \right)\right] + {1 \over c^2}
  a_A^j
  s_A^{ij} - {3 \over c^2} {d \over dt} \left(a_A^j m_A^{ij}\right)
  \nonumber \\
 & &  - {1 \over c^2} z_A^i \sum_l {1 \over l!}
  \left[l M_L^{A (1)} G_L^{A\prime}+\left(l+1\right)M_L^A G_L^{A (1)
  \prime}\right]
  - {1 \over c^2} \left(v_A^i z_A^j - \delta_{ij}
  {\bf v}_A.{\bf z}_A\right) R^j_{A a} \sum_l {1 \over l!} M_L^A
  G_{aL}^{A\prime}  \nonumber \\
 & &  - {1 \over c^2} \left(z_A^i z_A^j - {1 \over 2}
  \delta_{ij} z_A^2 \right) R^j_{A a} \sum_l {1 \over l!} {d \over
  dT_A} \left(M_L^A G_{aL}^{A\prime}\right)
 - {2 \over c^2} z_A^j R^i_{A [a} R^j_{A b]}
  \sum_l {1 \over l!} {d \over dT_A} \left(M_{aL}^A G_{bL}^{A\prime}
  \right) \biggr\} \nonumber \\
 & & + \sum_A \biggl\{ {d \over dT_A} \left(m^i_A\right)\left(1+{2
  \over c^2} G_A^{\prime\prime} \right) + {1 \over c^2} v_A^j {d
  \over dt}
  \left(m^j_A v_A^i\right) + {1 \over c^2} m^i_A {d \over dT_A}
  G_A^{\prime\prime} - {2 \over c^2} v_A^i m^j_A a_A^j\biggr\}
 \nonumber \\
 & &  + {\cal O}(4) \;. \label{twentyone}
\end{eqnarray}
\narrowtext
\noindent Again, the last sum of terms in curly brackets disappears
in the standard worldline gauge.

The LD-EIH form of the linear momentum is obtained by retaining only
the mass monopole terms
\begin{eqnarray}
 p^i_{LD-EIH} &=& \sum_A \biggl\{ M^A v_A^i \left[1+{1 \over 2c^2}
  \left(v_A^2 - \sum_{B \neq A} {GM^B \over r_{AB}}\right)\right]
  \nonumber \\
 & &  \quad\quad - {G \over 2c^2} \sum_{B \neq A} {G M^A M^B
  \over r_{AB}}\left({\bf n}_{AB}.{\bf v}_B\right) n^i_{AB}\biggr\}
 \nonumber \\ & & \quad\quad +{\cal O}(4) \; . \label{twentytwo}
\end{eqnarray}
The presence of spin dipoles (in the framework of the PD
model) results in the following additional term
\begin{equation}
 p^i_{PD} = p^i_{LD-EIH} + {1 \over c^2} \sum_A a^j_A s^{ij}_A +
 {\cal O}(4) \; , \label{twentythree}
\end{equation}
as mentioned in \cite{ds}.

As in the case of the total mass-energy, but with more work, one
can directly
check that the global-time derivative of Eq.\ (\ref{twentyone})
vanishes to the indicated accuracy.

Due to the incompatibilities between the DSX and the Fock approaches
mentioned above, it is not possible to compare directly our results
with those given by Fock \cite{fo} or Brumberg \cite{bru}.

\subsection{Angular momentum}

Damour and Iyer \cite{di1} and Ref.~\cite{dam3} have shown that the
treatment of the local individual spin of bodies, members of an
interacting $N$-body system, faces serious problems due to the
unavoidable intervention of nonlinear, interbody gravitational
effects. Mathematically, this complication manifests itself through
the occurrence of
``bad'' DSX moments $(N_L,P_L)$. It has been, however, possible to
reach
a successful formulation of the individual rotational laws of motion
through a carefully adjusted definition of the individual spin of
each body \cite{dam3}. In the following, we briefly address the
problem of
breaking up the total angular momentum into a sum of contributions
which make sense within the DSX framework.

The global angular momentum of the system reads (\cite{fo},
\cite{di1}, \cite{dam3})
\begin{eqnarray}
 s_i(t) &=& \epsilon_{ijk} \int d^3x\, x^j \biggl\{ \sigma^k \left[1+
  {4 \over c^2} w\right] \nonumber \\
 & & \qquad\;  - {\sigma \over c^2} \left[4 w^k + {1 \over 2}
  \partial_k \partial_t z(x,t)\right]\biggr\} + {\cal O}(4) \; ,
 \label{twentyfour}
\end{eqnarray}
where $z(x,t) = G \int d^3x^\prime \,\sigma(x^\prime,t) |{\bf x} -
{\bf x}^\prime |$. For an isolated system of arbitrary bodies it has
been previously shown that $s_i(t)$ is conserved at
the post-Newtonian level.

In order to preserve the post-Newtonian accuracy [i.e. modulo
${\cal O}(4)$] of the result,
one needs the transformation law of the global mass current
$\sigma^k(x)$
to local coordinate quantities with corresponding precision.
After some algebra one arrives at
\widetext
\begin{eqnarray}
 \sigma^i(x) &=& \left|{\partial X \over \partial x}
  \right|^{(3)} \biggl\{ v_A^i e^0_{A 0} \Sigma + e^i_{A a} \Sigma^a
  \nonumber \\
  & & \qquad\qquad\; + {1 \over c^2} \Bigl[ 2v_A^i G_{A}^{\prime
  \prime} + c^2
  e^{i (1)}_{A a} X^a - v_A^i A_{A a}X^a + R^i_{A a}\left({1 \over 2}
  A^{(1)}_{A a} X^2 - A^{(1)}_{A b} X^b X^a\right)\Bigr] \Sigma
 \nonumber \\
  & & \qquad\qquad\; +  {1 \over c^2} \Bigl[2R^i_{A a} G_A^{\prime
   \prime}
   + v_A^i V_A^a - R^i_{A a} A^b_{A} X^b + R^i_{A b} \partial_a
  \left({1 \over 2} A^b_{A} X^2 - A^c_{A} X^c X^b\right)\Bigr]
  \Sigma^a \nonumber \\
 & & \qquad\qquad\; - {1 \over c^2}\Bigl[v_A^i \partial_T\left(X^b
  V_A^b \Sigma\right)+ R^i_{A a} \partial_T\left( X^bV_A^b \Sigma^a
  \right)\Bigr] + {1 \over c^2} V_A^a R^i_{A b} \left[{\bf T}^{ab} -
  \delta_{ab} {\bf T}^{cc}\right] \biggr\} \nonumber \\
 & & + {\cal O}(4) \; , \label{twentyfive}
\end{eqnarray}
\narrowtext
\noindent (where $e^{i (1)}_{A a} = de^i_{A a} / dT_A$, etc.)
generalizing the formula $\sigma^i = v_A^i \Sigma + R^i_{A a}
\Sigma^a + {\cal O}(2)$ used throughout the series of papers
\cite{dam1,dam2,dam3,dam4} [Note the three-dimensional Jacobian,
as defined in Eq.\ (\ref{six}) above, in front of
Eq.\ (\ref{twentyfive})].

Inserting this relation into the defining integral (\ref{twentyfour})
and using the
method of Sec.II.A one obtains an expression of the form
\begin{equation}
 s_i(t) = \sum_A \left[ \Psi^A_i\left(M_L,S_L,G_L,H_L;P_L,N_L\right)
 + \theta_{1 i}^A + \theta_{2 i}^A\right] \; , \label{twentysix}
\end{equation}
where $\Psi^A_i$ is a function of both the good DSX moments
$(M_L,S_L,G_L,H_L)$ and the bad ones $(P_L,N_L)$, and where the
remaining two terms (which follow from the last term in
Eq.\ (\ref{twentyfive})) read
\begin{eqnarray*}
 c^2 \theta_{1 i}^A =& \epsilon_{ijk} z_A^j V_A^a R^k_{A b}
  \int_A d^3X_A \left[{\bf T}^{ab} - \delta_{ab} {\bf T}^{cc}\right]
  \; , \\
  c^2 \theta_{2 i}^A =& \epsilon_{ijk} R^j_{A a} V_A^b R^k_{A c}
  \int_A d^3X_A \,X^a\left[{\bf T}^{bc} - \delta_{bc} {\bf T}^{dd}
  \right] \; .
\end{eqnarray*}
Note that if we augment our list of ``bad'' moments by
including $Q^A_{ab} = \int_A d^3X_A\, {\bf T}^{ab}$ and $Q^A_{ab;c} =
\int_A d^3X_A\, {\bf T}^{ab}X_A^c$, we can express the total spin
$s_i(t)$
in terms of some individual moments of the $N$-bodies. We tried to
get rid of the non-dynamical moments $(P^A_L,N^A_L,Q^A_L)$ by using
the local conservation of energy-momentum $(\nabla_\alpha
{\bf T}^{\alpha
\beta}=0)$ to connect them to the dynamical moments $(M^A_L,S^A_L)$
and the tidal ones $(G^A_L,H^A_L)$. We did not succeed in doing so.
[In fact, such transformations introduced an undesirable dependence
upon the internal part $W^{\alpha +}_A(X)$ of the local gravitational
potential]. We thus hypothesize that the total spin $s_i$ of the
system
is not algebraically reducible to the DSX dynamical quantities.

\section{Conclusion}

The algebraic form of the global relativistic conservation laws has
been examined within the perspective of the DSX
post-Newtonian dynamics of an $N$-body system. We succeeded in
breaking up seven of these conservation laws (mass-energy,
center-of-mass quantity, and linear momentum) into a sum of
individual contributions involving only the basic dynamical quantities
of the DSX formalism. The angular momentum conservation law resisted,
however, our efforts.

\acknowledgments

D. V. finished this work while staying at the OCA/CERGA, Grasse
(France)
and being supported by an H. Poincar\'e fellowship. He is also
grateful to IHES, Bures sur Yvette (France) for its kind hospitality
and partial support.

\end{document}